\documentstyle[aps,multicol,epsfig]{revtex}
\def\be{\begin{eqnarray}}
\def\ee{\end{eqnarray}}

\renewcommand{\narrowtext}{\begin{multicols}{2} \global\columnwidth20.5pc}
\renewcommand{\widetext}{\end{multicols} \global\columnwidth42.5pc}

\def\inseps#1#2{\def\epsfsize##1##2{#2##1} \centerline{\epsfbox{#1}}}

\def\top#1{\vskip #1\begin{picture}(290,80)(80,500)\thinlines \put(65,500){\line( 1, 0){255}}\put(320,500){\line( 0, 1){
5}}\end{picture}}
\def\bottom#1{\vskip #1\begin{picture}(290,80)(80,500)\thinlines \put(330,500){\line( 1, 0){255}}\put(330,500){\line( 0, -1){
5}}\end{picture}}

\begin{document}
\title{Phase Separation Based on U(1) Slave-boson\\
Functional Integral Approach to the $t$-$J$ Model}
\author{Tae-Hyoung~Gimm$^1$ and Sung-Ho~Suck Salk$^{1,2}$}
\address{$^1$Department of Physics,
Pohang University of Science and
Technology, Pohang 790-784, Korea\\
$^2$Korea Institute of Advanced Studies, Seoul 130-012, Korea}
\date{29 September 1999}

\maketitle

\begin{abstract}
We investigate the phase diagram of phase separation
for the hole-doped two dimensional system of antiferromagnetically
correlated electrons
based on the U(1) slave-boson functional integral approach to the
$t$-$J$ model.
We show that the phase separation occurs for all values of
$J/t$, that is,
whether $0 < J/t < 1$ or $J/t \geq 1$ with $J$,
the Heisenberg coupling constant and $t$, the hopping strength.
This is consistent with other numerical studies of
hole-doped two dimensional antiferromagnets.
The phase separation in the physically interesting $J$ region,
$0 < J/t \lesssim 0.4$ is examined by introducing hole-hole
(holon-holon) repulsive interaction. We find from this study that
with high repulsive interaction between holes the phase separation
boundary tends to remain robust in this low $J$ region, while in the
high $J$ region, $J/t > 0.4$, the phase separation boundary tends to disappear.
\end{abstract}

\bigskip

\def\be{\begin{eqnarray}}
\def\ee{\end{eqnarray}}
\narrowtext
One of the most interesting observations in high-$T_c$ 
cuprates (superconductors) is the 
phase separation, which may play an important
role on superconductivity.
The phase separation results from 
a thermodynamic instability which arises from
the violation of the stability condition,
$K^{-1} = n^2\partial^2 e/\partial n^2 =
n^2\partial \mu/\partial n > 0$. Here
$K$ is the compressibility; 
$e$, the ground state energy per site;
$n$, the electron density; and $\mu$, the chemical
potential. 
Initially the phase separation instability was believed to inhibit
superconductivity. Recently it draws a 
great attention owing to its possible connection with
superconductivity \cite{castellani,emery2} based on
experimental observations \cite{tranquada,hunt}
in high-$T_c$ cuprate oxides.

We write the $t$-$J$ Hamiltonian for the study of
the hole doped systems of antiferromagnetically correlated electrons,
\be \label{tJ}
H && =
-t\sum_{\langle i,j \rangle\sigma}
\left (c^{\dagger}_{i\sigma}c_{j\sigma} + h.c.
 \right)  
+ J\sum_{\langle i,j \rangle}
\left ( {\bf S}_i\cdot {\bf S}_j - \frac{n_in_j}{4} \right ) \/,
\ee
with ${\bf S}_i =
 1/2c^{\dagger}_{i\sigma}{\bf \sigma}_{\alpha\beta}c_{i\beta}$ and
$n_i = \sum_\sigma c^{\dagger}_{i\sigma}c_{i\sigma}$
where $c^{\dagger}_{i\sigma}$ creates an electron
of spin $\sigma$ on site $i$. 
Earlier, using the $t$-$J$ model a possibility of 
phase separation in high-$T_c$ cuprates has been brought up 
by Emery et al. \cite{emery1}. They predicted the 
existence of phase separation at all possible values of
$J/t$, that is, $0 < J/t \leq 1$ or $J/t > 1$ where $J$ 
is the antiferromagnetic correlation strength and 
$t$, the hopping integral, including the case of
$J/t < 1$.
On the other hand, other numerical studies 
\cite{fehske,putikka,poilblanc,ogata,kohno} predicted the
existence of phase separation only for $J/t \geq 1$ where
$J$ value is unrealistic for the high $T_c$ cuprates of current interest.
Recently, from a Green function Monte Carlo study 
Hellberg and Manousakis \cite{hellberg} 
reported that the phase separation can occur for all values of $J$, in
agreement with the earlier exact diagonalization study of Emery et al \cite{emery1}.
In the present study,
by using the U(1) slave-boson functional integral method
\cite{kotliar,fukuyama,ioffe,lee,ubbens,gimm},
we obtain a phase diagram in the plane of electron 
density vs. $J/t$, by using the Maxwell construction 
\cite{emery1,bachelet}.

If violation of the stability condition $K^{-1} > 0$
occurs in the electron density range 
of $n_1 < n_e < n_2$, where $n_1$ is the 
electron density for a hole-rich phase and $n_2$,
the electron density for a hole-free phase,
the system is expected to separate into two
subsystems with electron densities $n_1$ and $n_2$ respectively.
Since we are interested in the hole-doped systems
of high $T_c$ cuprate oxides, 
the physics can be conveniently described
in terms of the hole density, $x = 1 - n_e$.
Thus we first examine
the ground state energy density
$e_h(x)$ of the hole doped system 
as a function of hole density $x$.
For the Maxwell's construction \cite{emery1,bachelet} to
treat a finite system, we consider 
a straight line 
which intercepts both the Heisenberg energy $e_H = e_h(0)$
at $x = 0$ and a curve given by
$e_h(x)$ at $x \neq 0$. The slope 
of the straight line at $x$ is then given by
\be  
e(x) = \frac{e_h(x) - e_H}{x} \/.
\ee
If a minimum of $e(x)$ is found at a hole density of $x=x_c$ 
(at which 
the straight line intercepts tangentially the curve of
$e_h(x)$),
phase separation is expected to exist below the 
onset (critical) density of $x = x_c$ \cite{emery1,bachelet}. 
As a result 
the energy density of the phase separated system is described by the the linear 
function with slope of $e(x_c)$
in the doping range of $0 < x < x_c$. 
Thus in this region
the system is stabilized with its energy lower than
that of the uniform phase, by forming a system composed of
two subsystems: one with
a hole-rich phase of the electron density of
$n_1 = 1 - x_c$ and the other with a hole-free phase of the
electron density of $n_2 = 1$.

In order to clarify how to compute the ground state energy
as a function of electron or hole density we briefly
discuss our earlier approach \cite{gimm} of the U(1) 
slave-boson representation of the $t$-$J$ Hamiltonian.
In this approach
we introduce
an additional contribution of hole-hole repulsion to 
the original $t$-$J$ Hamiltonian,
\be \label{H}
H && =
-t\sum_{\langle i,j \rangle\sigma}
\left (f^{\dagger}_{i\sigma}b_ib^{\dagger}_jf_{j\sigma} + h.c.
 \right)  
+ J\sum_{\langle i,j \rangle}
\left ( {\bf S}_i\cdot {\bf S}_j - \frac{n_in_j}{4} \right )  \nonumber \\
&& +V\sum_{\langle i,j \rangle}
b^{\dagger}_ib_ib^{\dagger}_jb_j
-\mu_0\sum_i\left (f^{\dagger}_{i\sigma}f_{i\sigma} - N_e \right )
\ee
with ${\bf S}_i =
 1/2f^{\dagger}_{i\sigma}{\bf \sigma}_{\alpha\beta}f_{i\beta}$ and
$n_i = \sum_\sigma f^{\dagger}_{i\sigma}f_{i\sigma}$.
$\mu_0$ is the chemical potential to fix the number of 
electron to $N_e$.
Here the
local constraint of single occupancy, $\sum_\sigma f^{\dagger}_{i\sigma}f_{i\sigma} +
 b^{\dagger}_ib_i = 1$ is assumed.
$f^{\dagger}_{i\sigma} (f_{i\sigma})$ is the spinon
creation (annihilation) operator and
$b_i (b^{\dagger}_i)$, the holon annihilation (creation)
 operator.
The nearest neighbor (NN) configuration of two
holes is energetically more favorable than 
other possible configurations. 
This is evident from the separate inspection of the two
attractive interaction
terms, $J {\bf S}_i\cdot {\bf S}_j$
and $-(J/4) n_in_j$. 
For the latter we write \cite{gimm},
\be \label{ninj}
&& -J\sum_{\langle i,j \rangle}\frac{n_in_j}{4}
= -\frac{J}{4}\sum_{\langle i,j \rangle}
\left \{ 1 - b^{\dagger}_ib_i - b^{\dagger}_jb_j
+ b^{\dagger}_ib^{\dagger}_jb_ib_j \right \} \nonumber \\
&&= -\frac{J}{2}\sum_{i\sigma}f^{\dagger}_{i\sigma}f_{i\sigma}
+\frac{J}{2}\sum_ib^{\dagger}_ib_i
- \frac{J}{4}\sum_{\langle i,j \rangle}
b^{\dagger}_ib^{\dagger}_jb_ib_j \/.
\ee
where the local constraint of  single occupancy is taken into
account.
The effective attraction between
the NN holes
arises 
from the last term of the equation above.
In view of the numerical finding of excessive large binding of
hole pairs \cite{gazza}, the hole-hole repulsive interaction
term (the third term in Eq.~(\ref{H})) is introduced.
Owing to its introduction, we are
now able to examine how phase diagram boundary 
is affected by the variation of hole-hole repulsion
interaction $V$. Such discussion will be made later.

As a result of the Hubbard-Stratonovich transformation 
\cite{ubbens} in Eq.~(\ref{H}),
the Heisenberg exchange and the hopping terms
are led to linearized terms involving the hopping order field
$\chi_{ji} = \langle {8t}/{3J}b^{\dagger}_jb_i 
+ f^{\dagger}_{j\sigma}f_{i\sigma} \rangle$ 
in association with the exchange interaction channel and 
the spinon singlet pairing order field
$\Delta^{{f}}_{ji}
 = \langle f_{j\uparrow}f_{i\downarrow}-f_{j\downarrow}f_{i\uparrow}
\rangle $ 
in association with the pairing channel.
The contribution of the direct (Hatree) channel is 
omitted based on 
the assumption of
paramagnetic states for each site, 
i.e., $\langle {\bf S}_i \rangle = 0$ 
\cite{ubbens}. Long-range antiferromagnetic fluctuations 
are thus ignored in this approach \cite{ubbens,gimm}.
The resulting effective Hamiltonian is then \cite{gimm},
\widetext
\top{-2.8cm}
\be \label{H00}
&& H = \sum_{\langle i,j \rangle}\frac{3J}{8} \left [ |\chi_{ji}|^2
+ |\Delta_{ji}^{{f}}|^2
 - \left (\frac{8t}{3J}b^{\dagger}_jb_i +
f^{\dagger}_{j\sigma}f_{i\sigma}\right )\chi_{ji}
- h.c.
- (f_{j\uparrow}f_{i\downarrow}-f_{j\downarrow}f_{i\uparrow})
{\Delta_{ji}^{f}}^{\ast} - h.c.  \right ] \nonumber \\
&& + \frac{8t^2}{3J}\sum_{\langle i,j \rangle}(b^{\dagger}_j
b_i)(b^{\dagger}_ib_j)
 -\sum_{\langle i,j \rangle}
(\frac{J}{4}-V) b^{\dagger}_ib^{\dagger}_jb_ib_j 
-(\mu_0-\frac{1}{4})\sum_if^{\dagger}_{i\sigma}f_{i\sigma}
+\frac{J}{2}\sum_ib^{\dagger}_{i}b_{i} \\
&&-i\sum_i \lambda_i(f^{\dagger}_{i\sigma}f_{i\sigma}
+b^{\dagger}_ib_i - 1 ) \/, \nonumber
\ee
\bottom{-2.8cm}
\narrowtext
\noindent
where a Lagrange multiplier field $\lambda_i$ is
introduced to impose the local
constraint of single occupancy for both the spinon and the holon.
$\lambda_i$ will be absorbed into the effective chemical
potential of holon and spinon in the mean field approaches 
\cite{ubbens}.
The quartic holon term 
(the second term in Eq.~(\ref{H00}) above)
$\frac{8t^2}{3J}\sum_{\langle i,j \rangle}(b^{\dagger}_j
b_i)(b^{\dagger}_ib_j)$
is repulsive \cite{baskaran}.
It is important to realize that this term involves nothing but the 
holon exchange interaction.
Thus allowing the holon exchange channel for
the quartic holon term (the second term in Eq.~(\ref{H00})),
we obtain
\widetext
\top{-2.8cm}
\be 
\label{key}
\frac{8t^2}{3J}\sum_{\langle i,j \rangle}(b^{\dagger}_j
b_i)(b^{\dagger}_ib_j) =
\frac{8t^2}{3J}\sum_{\langle i,j \rangle}
\left (\langle b^{\dagger}_jb_i \rangle
b^{\dagger}_ib_j  +
b^{\dagger}_jb_i
\langle b^{\dagger}_ib_j \rangle - \langle b^{\dagger}_jb_i\rangle
\langle b^{\dagger}_ib_j \rangle \right )\/.
\ee
We find from our numerical calculation of 
the Maxwell construction that the above holon exchange
channel also affects phase separation, by effectively
reducing the hopping (kinetic) energy.
We find that with the neglect of the contribution of 
Eq.~(\ref{key}) the phase separation does not
occur even at sufficiently low doping.

The effective holon attractive interaction term (when
$0 \leq V < J/4$ in the third term of Eq.~(\ref{H00}))
$-\sum_{\langle i,j \rangle}
(\frac{J}{4}-V) b^{\dagger}_ib^{\dagger}_jb_ib_j$ 
can be decomposed into terms involving the direct,
exchange, and pairing channels.
The Hubbard-Stratonovich transformations and 
Bogoliubov-Valantin transformation in the momentum-space 
are made. As described in Ref.~\cite{gimm},
the mean field free energy at hole doping rate $x$ is then.
\be \label{free}
&& F_{\rm MF}(\chi, \Delta^{{f}},
\Delta^{{b}})/N = 2{J}_{\rm eff}|\chi|^2
+ \frac{3J}{4}{|\Delta^{f}|}^2 
+ (\frac{J}{2}-2{V}){|\Delta^{b}|}^2  \\
&& - 2T\sum_{k}\ln\left[\cosh(\beta E^{{f}}_k/2)
\right ] 
+ T\sum_{k}\ln\left[\sinh(\beta E^{{b}}_k/2)
 \right ] 
 + (\frac{1}{2}+x)\mu^{b} - x\mu^{f} \/, \nonumber 
\ee
\narrowtext
\noindent
Here $E^{{f}}_k = \sqrt{(\epsilon^{f}_k - \mu^{f})^2
 + {\Delta_k^{{f}}}^2}$ is
the quasi-particle excitation energy
for spinons and
$E^{{b}}_k =
\sqrt{(\epsilon^{b}_k - \mu^{b})^2 -
{\Delta_k^{b}}^2}$
for holons, where $\mu^f$ and $\mu^b$ are the effective chemical
potentials of spinon and holon respectively and
$\epsilon^{f}_k = -\frac{3J}{4}\chi \gamma_k$,
$\epsilon^{b}_k = -{2}{t}_{\rm eff}\chi \gamma_k$ with
$\gamma_k \equiv \cos{k_x} + \cos{k_y}$.
The effective Heisenberg coupling constant is 
${J}_{\rm eff} = \frac{3J\eta^2/2 + 4\eta t + J/4
 -V}{(2\eta + 8t/(3J))^2}$ and the effective holon hopping strength is
${t}_{\rm eff} = \frac{2\eta t+J/4-V}{2\eta + 8t/(3J)}$, where
$\eta$ is the ratio of spinon and holon order parameter
$\eta = \frac{\langle f^{\dagger}_{i\uparrow}f_{j\uparrow} \rangle}
{\langle b^{\dagger}_ib_j \rangle}$ \cite{gimm}.

From the minimization of the mean field free energy 
 with respect to the scalar fields,
$\chi, \Delta^{f},$ and $\Delta^{b}$,
we determine the ground state energy density of the hole-doped
system per site, $e_h(x)
= \lim_{T \rightarrow 0}F_{\rm MF}(\chi, \Delta^{{f}},
\Delta^{{b}})/N$ as a function of hole density $x$.
The critical hole density is actually found from
the Maxwell construction by introducing 
$e(x) = (e_h(x) - e_H)/x$. 
In Fig.~1 we display the Maxwell's construction for
$J = 2.5t$ and $V = 0$. The inset is the Green function Monte Carlos 
calculation of the $t$-$J$ model by Hellberg and Manousakis 
\cite{hellberg}. 
Quantitative disagreement exists between our U(1)
slave-boson functional integral approach and the
Green function Monte Carlo calculation. However
we find that there exists a minimum of $e(x)$ at the critical
hole density of $x_c \sim 0.72$  which is close to the 
value of Hellberg and Manousakis \cite{hellberg},
$x_c \sim 0.7$, 
below which phase separation occurs.

In Fig.~2 we display a predicted phase diagram in the plane of
the unitless Heisenberg exchange coupling strength, $J/t$ 
and the electron density, $n = 1 - x$ for various 
values of NN hole-hole (holon-holon) repulsion energy $V$.
The phase separation was predicted to occur for $J/t \leq 1$ and 
$J/t > 1$.
This prediction
is consistent with other numerical studies \cite{emery1,hellberg}.
Green function Monte Carlo results \cite{hellberg} 
(solid line) up to the $28 \times 28$ square lattice and 
and the exact diagonalization result \cite{emery1} (stars)
with a $4 \times 4$ lattice a
are displayed for comparison 
with our results 
(solid circles for $V = 0$) with 
\begin{figure}[b]
\inseps{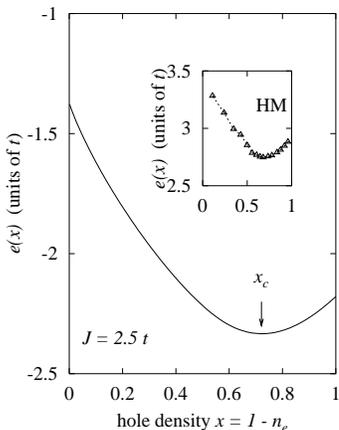}{0.66}
\caption{\label{fig1}
Maxwell's construction: $e(x)$ vs. $x$ for
$J = 0.1t$, $V=0$.
The Green function Monte Carlo result [11] in the inset is denoted by HM.}
\end{figure}
\noindent
\begin{figure}[b]
\inseps{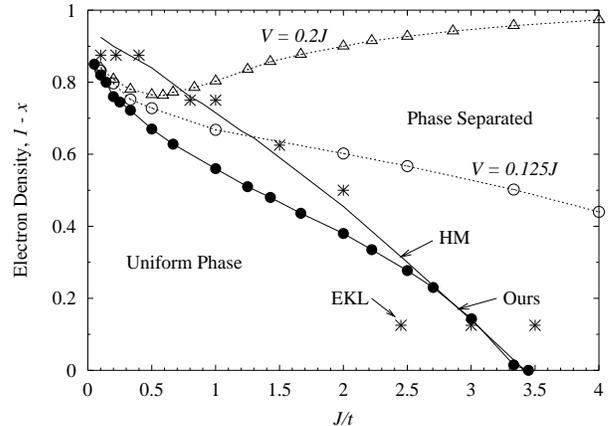}{0.66}
\caption{\label{fig2}
Phase separation for the hole-doped systems
of antiferromagnetically correlated electron
in the plane of Heisenberg coupling
strength, $J/t$ and the electron density, $n = 1 - x$.
The solid line denoted by HM is the Monte Carlo prediction
(Ref. [11]); and the stars denoted by EKL,
the result of Emery et al (Ref. [5]);
and the solid circles are our computed results
for $V=0$.
The phase separation boundaries for
$V = 0.125J$ (open circles) and
$V = 0.2J$ (open triangles) are also displayed.
The critical hole doping density $x_c$ is seen to
decrease with
the increase of $V$ in the region of large $J/t$.}
\end{figure}
\noindent
$100 \times 100$ lattice \cite{finite}.
Despite some numerical differences, interestingly all of
these methods yield nearly the same critical $J_c \sim 3.4t$,
above which the hole-rich phase contains no spins, as shown in
the Fig.~2.
In the small $J/t$ limit, the phase separation is expected
to occur as a result of relative increase in kinetic 
energy (compared to the Heisenberg interaction energy $J$) which
promotes relatively easier hopping of holon (holes) from site
to site, thus avoiding antiferromagnetic spinon (spin) frustrations
to lower the energy of the system and creating a hole-rich region.
On the other hand, in the large $J$ limit the phase separation
occurs owing to the Heisenberg interaction coupling which
promotes the antiferromagnetic order by inhibiting the 
occurrence of holes in the region of the antiferromagnetic
phase \cite{emery1}.

We now explore the effects of hole-hole (holon-holon)
repulsive interaction $V$ on the phase separation boundary.
The uniform phase is expected to be more favorable
owing to the enhanced difficulty of hole pairing with the increase
of $V$. The critical doping density in the physically interesting 
$J$ region, $J/t \lesssim 0.4$ is predicted to be relatively
insensitive to the variation of $V$ compared to the high $J$ limit,
as shown in Fig.~2. We observe that the critical
hole doping density $x_c$ quickly decrease beyond the large
$J$ region of $J/t > 0.4$ as the holon-holon (hole-hole) repulsion
energy $V$ increases.
For the case of the large $V$ limit ($V \sim 0.25J$),
the uniform phase occurs with a small critical
hole density, indicating the phase separation is not
likely to occur.
As shown in Fig.~2, we note the persistence of
phase separation in the region of small $J$ and 
the propensity of gradual disappearance of phase
separation in the region of large $J/t$ at this high limit of
the hole-hole repulsive interaction. Such
persistence of phase separation at small $J/t$ despite the 
increase of $V$ is attributed to the effective
increase of the kinetic energy of holons, to avoid
the frustration of antiferromagnetic spinons \cite{emery1}.

In the present study we investigated 
the phase diagram involving phase separation
based on the U(1) slave-boson functional integral approach to the
$t$-$J$ model.
We find that the phase separation occurs in the region of
low hole doping for all possible 
values of Heisenberg coupling constant $J$, that is,
whether $J/t < 1$ or $J/t \geq 1$ 
with an upper bound of $J/t$ (at $J \simeq 4.2t$).
This observation is consistent with other numerical studies of
hole-doped two dimensional antiferromagnets.

\widetext

\begin{references}
\bibitem{finite} Our numerical results are based
on Eq.~(\ref{free}) at $T = 0$ limit which
is expressed in momentum space. Numerical 
convergence was achieved for a $100 \times 100$ square
lattice, which yielded a good convergence.
\end{references}
\end{document}